\documentclass[10pt,letterpaper,twocolumn]{article} 

\usepackage{OL}
\usepackage{hyperref}
\usepackage{amsmath}

\begin{document}

\twocolumn[ 

\title{Quartic Kerr solitons}


\author{Hossein Taheri$^{1,\dagger}$ and Andrey B. Matsko$^2$}
\affiliation{$^1$Department of Electrical and Computer Engineering, University of California at Riverside, 900 University Ave., Riverside, CA 92521}
\vskip-8pt
\affiliation{$^2$OEwaves Inc., 465 North Halstead Street, Suite 140, Pasadena, California 91107, USA}
\email{$^\dagger$Corresponding author: hossein.taheri@ucr.edu}


\begin{abstract}
Solitons, ubiquitous in nonlinear sciences, are wavepackets which maintain their characteristic shape upon propagation. In optics, they have been observed and extensively studied in optical fibers. The spontaneous generation of a dissipative Kerr soliton (DKS) train in an optical microresonator pumped with continuous wave (CW) coherent light has placed solitons at the heart of optical frequency comb research in recent years. The commonly observed soliton has a ``sech''-shaped envelope resulting from resonator cubic nonlinearity balanced by its quadratic anomalous group velocity dispersion (GVD). Here we exploit the Lagrangian variational method to show that CW pumping of a Kerr microresonator featuring quartic GVD forms a pure quartic soliton (PQS) with Gaussian envelope. We find analytical expressions for pulse parameters in terms of experimentally relevant quantities and derive an area theorem. Predictions of the analytical calculations are validated with extensive numerical simulations. The broader bandwidth and flatter spectral envelope of a PQS, compared to a DKS of the same pulse width and peak power, make it superior for applications requiring small line-to-line power variation in frequency comb harmonics.
\end{abstract} 


 ] 

\noindent Solitons are particle-like shape-conserving wavepackets observed in various nonlinear systems \cite{drazin1989solitons}. Generally, dispersion and nonlinearity of the propagation medium dictate the pulse shape \cite{grelu2012dissipative}. For instance, balance of self-focusing cubic nonlinearity and anomalous quadratic group velocity dispersion (GVD) creates ``$\text{sech}$''-shaped bright solitons in optical fibers \cite{grelu2012dissipative}. Soliton properties can be modified through engineering dispersion and nonlinearity \cite{agrawal2013NLfiber}. This engineering task is difficult in standard optical fibers, although it has been pursued more effectively in microstructured and photonic crystal (PC) fibers \cite{russell2003pcf}. Additionally, expensive high-power mode-locked lasers (MLLs) are needed for fiber soliton generation. Recently emerged dissipative Kerr solitons (DKSs), on the other hand, have opened a new chapter in nonlinear optics by combining small footprint, reduced cost, unprecedented possibility in dispersion engineering, and reduction in required power for soliton generation \cite{kippenberg2018science}. The spontaneous generation of stable DKS trains has incentivized development of miniaturized sources of ultrashort optical pulses and, associated with them, broadband frequency combs \cite{herr2014temporal} for myriad applications in recent years, most notably for high-spectral-purity microwave and radio-frequency signal generation \cite{liang2015high}, dual-comb spectroscopy \cite{vahala2016dcs}, high-capacity optical communication \cite{koos2017opticalcomm}, light detection and ranging (LiDAR) \cite{koos2018ranging}, search for exoplanets \cite{herr2019astrocomb}, and timekeeping \cite{papp2018clock}. Microresonator-based DKS generation has proven a prolific and fast-rising area of nonlinear photonic technology.

Dispersion profile shaping through resonator morphology engineering is one of the most empowering features of nonlinear optics in microresonators \cite{okawachi2014BWshaping, li2017stably}. While standard nonlinear Schrodinger equation (NLSE) solitons hinge on nonlinearity offset by second-order dispersion, considering higher-order dispersion is inevitable when pulse spectral bandwidth increases \cite{agrawal2013NLfiber}. Solitons perturbed by third- and fourth-order dispersion (TOD and FOD) can travel stably \cite{taheri2016highorderdisp, milian2014soliton} in the company of a Cherenkov dispersive wave \cite{akhmediev1995cherenkov, taheri2017latticetrap}. Intentional inclusion of TOD and FOD has been employed to increase bandwidth of microcombs and simplify application of self-referencing stabilization techniques \cite{brasch2016photonic, li2017stably}. In what follows we focus on Kerr solitons when FOD is the leading resonator dispersion term. 

Stable ``quartic solitons'' were reported in early 1990s in optical fibers \cite{karlsson1994fod} and more recently in integrated photonic slot waveguides (WGs) \cite{biancalana2013quartic} when pulse center frequency is at a GVD local extremum with significant FOD contribution. While TOD was negligible in these systems, quadratic dispersion was still significant. More recently, quartic solitons were demonstrated experimentally in PC WGs featuring pure FOD \cite{blanco2016purequartic}. This ``pure quartic soliton'' (PQS) was shown to have a Gaussian, rather than hyperbolic secant, envelope, hence having a smaller width than an NLSE soliton having the same peak power \cite{saleh2007fundamentals}. The broader frequency coverage and flatter spectral envelope of a PQS microcomb, compared to a DKS, make it more attractive for applications demanding small harmonic-to-harmonic power variation, e.g., spectroscopy and optical communication.

In this Letter, we take a variational approach to show that under certain conditions a PQS train can be generated in a Kerr-nonlinear resonator with pure quartic modal dispersion. Unlike the recent demonstration in PC WGs, where MLL pulses shaped to the transform limit were utilized \cite{blanco2016purequartic}, microcomb PQSs are driven by CW lasers with modest power. We find analytical expressions relating PQS parameters to resonator and pump properties. These relations are validated by numerical simulations based on direct integration of a modified Lugiato-Lefever equation (LLE). We also derive an area theorem linking PQS energy to its temporal width and note that while peak energy is inversely proportional to pulse width for a DKS, it scales as inverse width cubed for a PQS. This scaling rule suggests that a short temporal PQS will have significantly larger peak power compared to a DKS.

\begin{figure}[tb]
\centering
\includegraphics[width=0.45\textwidth]{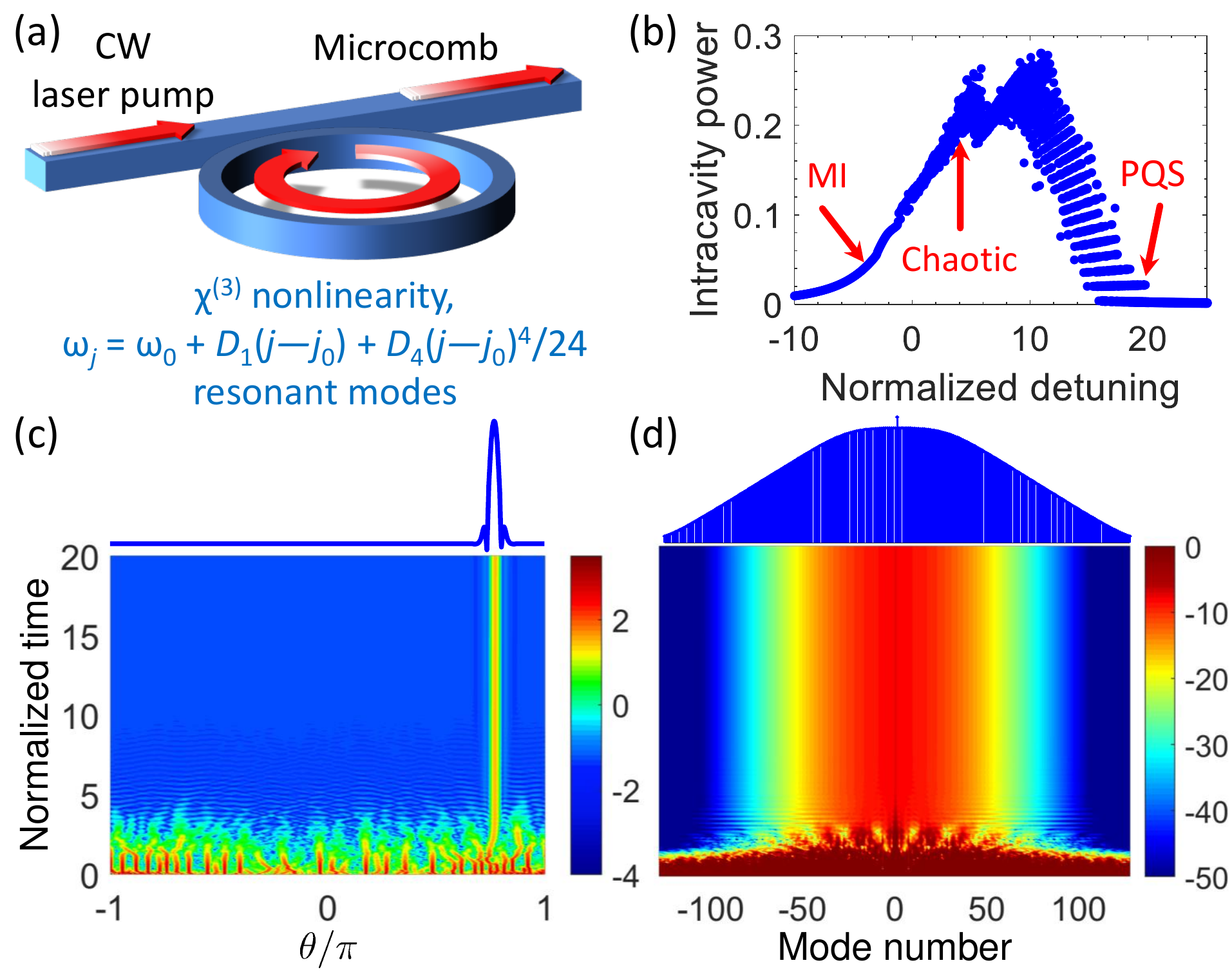}
\caption{(a) Schematic of a Kerr-nonlinear microresonator with pure quartic modal dispersion pumped by a CW laser via an access WG. (b) Microcomb power as a function of detuning, showing regions characterized by stable MI, chaotic behavior, and PQS formation. The step-like profile shows stable microcombs with different numbers of PQS peaks. The arrow labeled ``PQS'' refers to a single-PQS branch. (c) Temporal evolution of the field envelope upon hard excitation by a random high energy initial state creating a single-peak PQS, with final pulse power profile (in logarithmic scale) shown on top. (d) Temporal evolution of the frequency spectrum for (c), with final comb frequency spectrum (in dB) shown on top.
}
\label{spect}
\end{figure}

\begin{figure}[tb]
  \centering
  \includegraphics[width=0.45\textwidth]{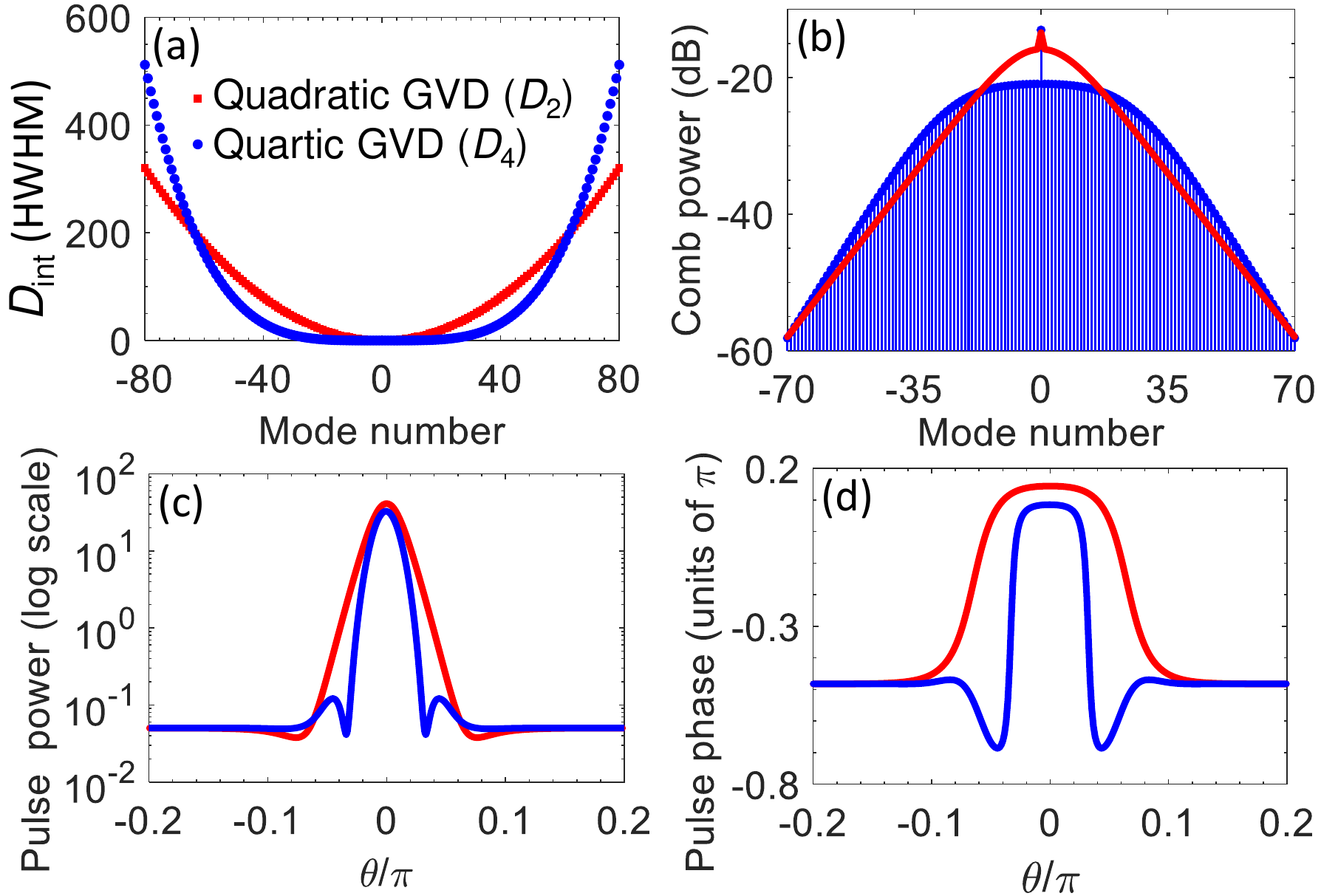}
\caption{Comparison of (a) pure quartic (blue dots) and pure quadratic (red squares) residual dispersion, (b) PQS (blue spikes) and DKS (red curve) microcomb spectra, (c) PQS (blue) and DKS (red) pulse shapes (power, in log scale), and (d) PQS (blue) and DKS (red) pulse temporal phase profiles.
} \label{comparisons}
\end{figure}

The schematic of Fig.~\ref{spect}(a) illustrates a microresonator (here a ring resonator) driven by a CW pump through an access WG (e.g., an integrated ridge WG). The resonator is characterized with Kerr nonlinearity (susceptibility $\smash{\chi^{(3)}>0}$) and pure quartic modal dispersion; its resonant mode frequencies $\omega_j$ can be Taylor expanded as
\begin{equation}
\omega_j =  \omega_{j_0} + D_1 (j - j_0) + D_4 (j - j_0)^4/4! \label{disp}.
\end{equation}
Here the center frequency $\omega_{j_0}$ with azimuthal mode number $j_0$ signifies the mode closest to the pump and $D_k = \partial^k\omega_j/\partial j^k$ for integer $k$ are coefficient in the series expansion of the modal frequencies in terms of the mode numbers $j$ at $j_0$. $D_1$ is resonator free spectral range (FSR) at $\omega_{j_0}$ and $D_4$ is the FOD coefficient. (See Fig.~\ref{comparisons}(a) for a comparison of quadratic and pure quartic GVD, where $D_{\text{int}} = \omega_j - \omega_{j_0} - D_1 (j - j_0)$ is the integrated or residual dispersion parameter and is expressed in units of resonator half-width at half-maximum (HWHM) bandwidth $\Delta\omega/2$.) In terms of WG dispersion \cite{agrawal2013NLfiber} (in the case of a ring or fiber-based resonator), $D_2 = D_3 = 0$ translates into vanishing $\beta_2$ and $\beta_3$ GVD parameters, and $D_4 = -\beta_4 v_{\text{g}}^5/R^4$, $v_{\text{g}}$ being WG group velocity and $R$ ring radius \cite{taheri2017thesis}. The resonator is considered over-coupled and the modal bandwidth $\Delta\omega$ depends on its intrinsic quality factor (Q) and loading. Microcomb formation in this structure is governed by the LLE \cite{matsko2011modelocked, chembo2013spatiotemporal, coen2013modeling} modified to include pure FOD \cite{taheri2016highorderdisp},
\begin{equation}
\frac{\partial\mathcal{A}}{\partial t} = \big( -\frac{\Delta\omega}{2} -\text{i}\sigma + \text{i}\frac{D_4}{4!} \frac{\partial^4}{\partial\theta^4} -\text{i} g \left|\mathcal{A}\right|^2 \big) \mathcal{A} + F. \label{LLE}
\end{equation}
$\mathcal{A}$ is the temporal intracavity field envelope, $t$ is the slow time, $\sigma = \omega_{\text{p}} - \omega_{j_0}$ is the detuning of the pump frequency $\omega_{\text{p}}$ from the pumped resonance, $\theta = v_{\text{g}} T/R$ (modulo $2\pi$) is the azimuthal angle (proportional to the fast time $T$), $g$ is the Kerr-induced single photon frequency shift, and $|F|^2$ quantifies pump power. Normalized quantities as follows are used for numerical simulations: field envelope $\psi = \mathcal{A}^*\sqrt{2g/\Delta\omega}$, time $\tau = t\Delta\omega/2$, detuning $\alpha = -2\sigma / \Delta\omega$, dispersion $d_4 = -2D_4/\Delta\omega$, and pump power $f = F\sqrt{8g/\Delta\omega^3}$.

Solving Eq.~(\ref{LLE}) using the split-split Fourier transform (SSFT) method \cite{agrawal2013NLfiber} for  $\alpha = 19$, $\smash{d_4 = -3\times 10^{-4}}$, and $\smash{f = \sqrt{20}}$ we find that a family of stable PQSs are supported by the resonator. Intracavity power versus normalized frequency detuning, depicted in Fig.~\ref{spect}(b), shows a familiar behavior typical of soliton microcomb formation through pump frequency scan \cite{herr2014temporal}. Modulational instabilitiy (MI) is observed when the pump is blue-detuned with respect to the cavity mode ($\omega_\text{p} > \omega_{j_0}$) and total comb power increases uniformly with normalized detuning. As pump frequency is further tuned, the system crosses a chaotic region before PQS pulses are formed in the red-detuned regime. Stable pulsed states are characterized by a multi-stable step-like profile where successive power steps signify multi-soliton states having different number of pulses. The temporal evolution of a single-PQS state and its frequency spectrum is shown in Figs.~\ref{spect}(c) and (d), respectively. This PQS is formed through hard excitation with a random high-energy initial comb. Our numerical simulations show that the conversion efficiency of PQS microcombs, like that of quadratic-quartic solitons \cite{taheri2016highorderdisp}, increases with increasing $D_4$.

\begin{figure}[tb]
  \centering
  \includegraphics[width=0.45\textwidth]{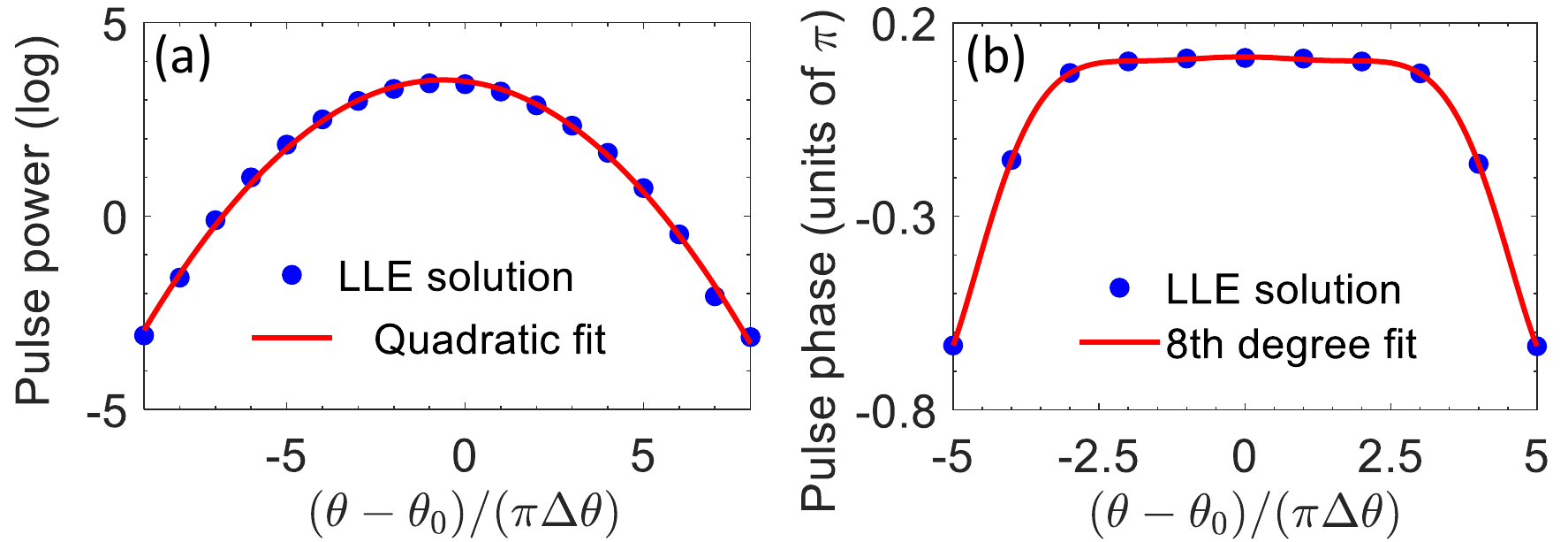}
\caption{Polynomial curve fitting (red) for (a) power (logarithmic scale) and (b) temporal phase of the final pulse shown on the top in Fig.~\ref{spect}(c). $\theta_0$ is pulse peak location and $\Delta\theta = 2\pi/265$ is the discretization step along the $\theta$-axis used in SSFT integration of Eq.~\ref{LLE}.
} \label{fits}
\end{figure}

A comparison between DKS and PQS microcombs shows that when pulse peak power is the same, PQS possesses a broader spectrum featuring a very flat envelope, Fig.~\ref{comparisons}(b), and the wings of its intensity profile demonstrate faster roll-off, Fig.~\ref{comparisons}(c). Also, the PQS temporal phase, Fig.~\ref{comparisons}(d), has a much flatter profile close to the pulse peak.
The main lobe of the PQS in the logarithmic-scale pulse shape of Fig.~\ref{comparisons}(c) (blue curve) is almost parabolic and curve fitting confirms its approximately Gaussian profile, Fig.~\ref{fits}(a). It is also seen that the flat PQS temporal phase near the pulse peak $\theta_0$ is closely fitted by an eighth-degree curve proportional to $(\theta-\theta_0)^8$ suggesting that the pulse is essentially chirp-free, Fig.~\ref{fits}(b).

Foregoing curve fitting results warrant adopting the Lagrangian variation method with a Gaussian ansatz to find pulse parameters such as peak power and width. We note that the analytic pulse shape found for quadratic-quartic NLSE solitons \cite{karlsson1994fod} simplifies to a trivial flat solution for vanishing quadratic dispersoin coefficient. The Lagrangian density (LD), $\mathcal{L}$, for a modified LLE with quartic dispersion term depends not only on the first derivatives $\mathcal{A}_t = \partial\mathcal{A}/\partial t$ and $\mathcal{A}_\theta = \partial\mathcal{A}/\partial\theta$ of the pulse shape $\mathcal{A}$, but also on its second- (or higher-) order derivatives. Therefore, the variation definition used for the standard LLE \cite{matsko2013timing, herr2014temporal} should be modified to accommodate the dispersion term in Eq.~\ref{LLE} \cite{gelfand2000cov, courant1962methods},
\begin{equation}
\frac{\delta\mathcal{L}}{\delta\mathcal{A}} = \frac{\partial\mathcal{L}}{\partial\mathcal{A}}-\frac{\partial}{\partial t} \Big[ \frac{\partial\mathcal{L}}{\partial\mathcal{A}_t} \Big] - \frac{\partial}{\partial\theta} \Big[ \frac{\partial\mathcal{L}}{\partial\mathcal{A}_\theta} \Big] + \frac{\partial^2}{\partial\theta^2} \Big[ \frac{\partial\mathcal{L}}{\partial\mathcal{A}_{\theta\theta}} \Big] = \mathcal{R}^*. \label{VA}
\end{equation}
Here, $\mathcal{R} = \text{i}(-\Delta\omega\mathcal{A}/2+F)$ is the perturbation. It may be readily verified, e.g., by the direct substitution in Eq.~\ref{VA}, that the following LD (at least $C^1$ in $\mathcal{A}$ and its derivatives, except possibly on a set of measure zero) renders the desired governing LLE of Eq.~\ref{LLE},
\begin{equation}
\mathcal{L} = \frac{\text{i}}{2} \Big( \mathcal{A}^* \frac{\partial\mathcal{A}}{\partial t} -  \mathcal{A} \frac{\partial\mathcal{A^*}}{\partial t} \Big) + \frac{D_4}{4!} \left|\frac{\partial^2 \mathcal{A}}{\partial\theta^2}\right|^2 - \frac{g}{2} \left|\mathcal{A}\right|^4 - \sigma \left|\mathcal{A}\right|^2. \label{LD}
\end{equation}
We select a Gaussian ansatz
\begin{equation}
\mathcal{A} = A \exp \left[ \text{i}\phi - \text{i}h(\theta-\theta_0) - (1 + \text{i}C)(\theta-\theta_0)^2/2w^2 \right] \label{ansatz}
\end{equation}
with six unknown parameters, i.e., pulse amplitude, phase, delay, width, frequency and chirp ($A$, $\phi$, $\theta_0$, $w$, $h$ and $C$, respectively). Calculating the Lagrangian $\smash{L = \int_{-\pi}^\pi \mathcal{A} \, \text{d}\theta}$ and using the Euler-Lagrange equations including perturbation,
\begin{equation}
\frac{\partial L}{\partial q} - \frac{\partial}{\partial t} \Big( \frac{\partial L}{\partial \dot{q}} \Big) = \int_{-\pi}^\pi \text{d}\theta \, \Big( \mathcal{R} \frac{\partial\mathcal{A}^*}{\partial q} + \mathcal{R^*} \frac{\partial\mathcal{A}}{\partial q} \Big) \label{EL},
\end{equation}
(where the generalized coordinate $q$ refers to $A, \phi, \theta_0, w, h,$ and $C$, and the overdot denotes time derivative) a set of six equations can be derived for the temporal evolution of pulse parameters. These equations cannot be solved analytically, but with clues found from curve fitting, Fig.~\ref{comparisons}(d) and Fig.~\ref{fits}(b), $h$ and $C$ can be set to zero, leading to significant simplification. The temporal evolution of pulse delay then follows $\text{d}\theta_0/\text{d}t = 0$, meaning that once a PQS is formed, it remains stationary; this agrees with the numerical integration result as well as the circular symmetry of Eq.~\ref{LLE}. Steady state pulse amplitude and width are found to be
\begin{equation}
A = \sqrt{\frac{-8\sqrt{2}}{7} \cdot \frac{\sigma}{g}} \quad \mathrm{and} \quad
w = \frac{1}{2} \sqrt[4]{\frac{-7}{2} \cdot \frac{D_4}{\sigma}}, \label{PAW}
\end{equation}
and $\smash{\phi = \cos^{-1}\Big[ (\Delta\omega/F) \sqrt{(-\sqrt{2}/7) \cdot (\sigma/g)} \Big]}$.

Equations \ref{PAW} predict a linear relationship between pulse peak power and pump-resonance detuning ($A^2$ vs. $\sigma$), and between the fourth power of pulse width and FOD coefficient as well as inverse detuning (i.e., $w^4$ vs. $D_4$, and $w^4$ vs. $\sigma^{-1}$). The numerical values of pulse peak power and width found as a function of detuning and FOD parameters using numerical integratoin of Eq.~\ref{LLE} over PQS existence range [c.f., Fig.~\ref{spect}(b)] agree remarkably well with these predictions.

\begin{figure}[tb]
  \centering
  \includegraphics[width=0.45\textwidth]{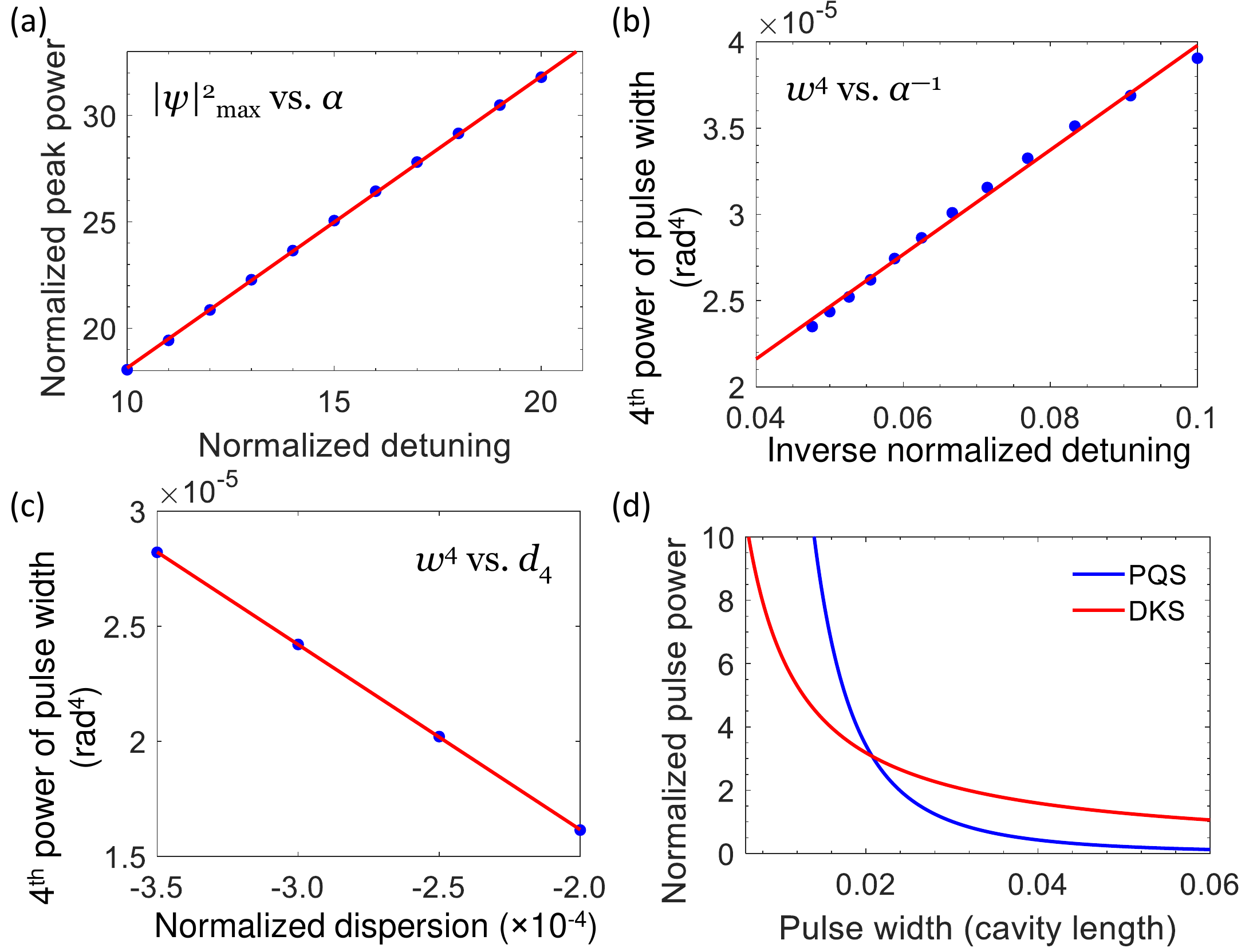}
\caption{Pulse parameters measured using SSFT integration of Eq.~\ref{LLE} (blue dots) showing that the relationship between (a) peak pulse power vs. detuning, (b) 4th power of pulse width vs. reciprocal of pump-resonance detuning ($w^4$ vs. $\alpha^{-1}$), and (c) 4th power of pulse width vs. FOD coefficient ($w^4$ vs. $d_4$) are linear (red), as predicted by the variational approach, Eqs.~\ref{PAW}. (d) Comparison of area theorems for PQS (blue) and DKS (red).
} \label{parameters}
\end{figure}

Pulse peak power and width expressions of Eqs.~\ref{PAW} can be used to find an area theorem (scaling rule) linking the energy $E = A^2 w $ and width of the PQS,
\begin{equation}
E_{\text{PQS}} = \frac{\sqrt{2} D_4}{4 g w^3}.
\end{equation}
It is seen that, compared to $E_{\text{DKS}} = D_2/(g w)$ for a DKS \cite{matsko2013timing}, PQS energy rises more rapidly with decreasing pulse width, particularly for shorter pulses, Fig.~\ref{parameters}(d).

In conclusion, extending the parallelism between nonlinear phenomena in optical fibers and microresonators, we have shown, numerically and analytically, that a class of bright pulses with Gaussian envelope (PQSs) arise from the interaction of pure quartic modal dispersion and Kerr nonlinearity in high-Q optical microresonators. In contrast to PQSs observed in PC WGs, pumping with a pulsed laser is not required. Analytic expressions for PQS parameters in terms of resonator design and experimentally tunable quantities are derived using the perturbative Lagrangian variational method, and are validated by numerical simulations based on direct integration of a modified LLE. While DKS energy increases linearly with decreasing pulse width, the area theorem derived here shows that this relationship is cubic for a PQS. The broad bandwidth and flat spectral envelope of a PQS microcomb make it advantageous for applications calling for a coherent frequency comb with small line-to-line power variation.

We thank Qing Li of Carnegie Mellon University for helpful comments.


%
\bibliographystyle{osajnl}
\bibliography{references}

\end{document}